\documentclass[a4paper, amsfonts, amssymb, amsmath, preprint, showkeys, nofootinbib, twoside]{revtex4-1}
\usepackage{graphicx} 
\usepackage{amsmath}
\usepackage{hyperref}
\usepackage{algorithm}
\usepackage{algorithmic}
\usepackage{placeins}

\usepackage{todonotes}

\begin{document}

\title{Supply Chain Due Diligence Risk Assessment for the EU: A Network Approach to estimate expected effectiveness of the planned EU directive}
\author{Jan Hurt$^{1,2}$, Katharina Ledebur$^{1,2}$, Birgit Meyer$^{1,3}$, Klaus Friesenbichler$^{1,3}$, \\ Markus Gerschberger$^{1,4}$, Stefan Thurner$^{1,2,5,6}$, Peter Klimek$^{1,2,5}$}
\email[peter.klimek@ascii.ac.at]{}
\affiliation{$^1$Supply Chain Intelligence Institute Austria, Josefstädter Strasse 39, A-1080 Vienna, Austria
\\
$^2$Complexity Science Hub Vienna, Josefstädter Strasse 39, A-1080 Vienna, Austria,
\\
$^3$Austrian Institute of Economic Research. Arsenal Objekt 20, A-1030 Vienna,
\\
$^4$Josef Ressel Centre for Real-Time Value Network Visibility, Logistikum, FHOÖ, Wehrgrabengasse 1-3, A-4400 Steyr.
\\
$^5$Section for Science of Complex Systems, CeDAS, Medical University of Vienna, Spitalgasse 23, A-1090 Vienna, Austria,
\\
$^6$Santa Fe Institute, 1399 Hyde Park Road, Santa Fe, NM 87501, USA.}

\begin{abstract}
Globalization has had undesirable effects on the labor standards embedded in the products we consume. This paper proposes an ex-ante evaluation of supply chain due diligence regulations, such as the EU Corporate Sustainable Due Diligence Directive (CSDDD). 
We construct a full-scale network model derived from structural business statistics of 30 million EU firms to quantify the likelihood of links to firms potentially involved in human rights abuses in the European supply chain.
The 900 million supply links of these firms are modeled in a way that is consistent with multiregional input-output data, EU import data, and stylized facts of firm-level production networks.
We find that this network exhibits a small world effect with three degrees of separation, meaning that most firms are no more than three steps away from each other in the network.
Consequently we find that about 8.5\% of EU companies are at risk of having child or forced labor in the first tier of their supply chains, about 82.4\% are likely to have such offenders at the second tier and more than 99.1\% have such offenders at the third tier.
We also profile companies by country, sector, and size for the likelihood of having human rights violations or child and forced labor violations at a given tier in their supply chain, revealing considerable heterogeneity across EU companies.
Our results show that supply chain due diligence regulations that focus on monitoring individual buyer-supplier links, as currently proposed in the CSDDD, are likely to be ineffective due to a high degree of redundancy and the fact that individual company value chains cannot be properly isolated from the global supply network.
Rather, to maximize cost-effectiveness without compromising due diligence coverage, we suggest that regulations should focus on monitoring individual suppliers.
\end{abstract}

\maketitle

\section{Introduction}
In recent decades, production and trade across value chains have become increasingly globalised. According to Eurostat, 48.8\% of all goods exported from the EU in 2022 were intermediate goods \cite{eurostat2023trade}. This mirrors the observation that value chains have become internationally fragmented. Parts and components are transported multiple times across borders at various stages of the manufacturing process \cite{timmer2014slicing}. The globalisation of value chains is elevating awareness of potential due diligence issues associated with human rights (HR) violations along  supply chains \cite{smit2021human}.
Firms are increasingly held accountable for the social and environmental impacts of not only their own activities, but also for the actions of their suppliers and their suppliers' suppliers, and so on \cite{GONG201988}.

Along with international trade, firms have partly relocated production processes from developed to less developed countries where labour costs are typically lower.
This has led to a more internationalized division of labor and a more cost efficient allocation of production. However, labor regulations are typically less effective in less developed countries. This has not only raised concerns about a level playing field for firms, as promoted by the World Trade Organization, but also about modern slavery, including child and forced labour \cite{gold2015modern}.
As of 2021, an estimated 27.6 million people worldwide, or 3.5 out of every thousand individuals, encountered  situations of forced labour on any given day. This marked an increase of 2.7 million individuals compared to estimates of 2016. \cite{international2022global}
Forced labour pervades all sectors of the private economy, encompassing manufacturing, construction, agriculture, and mining \cite{hofmann2018conflict}.
The NGO Jewish World Watch estimates that since 2017, more than two million Uyghurs have been forcibly relocated to labour camps, which are used by more than 2,000 multinational companies in their supply chain \cite{JWW}.
According to the International Labour Organization, an estimated 3.3 million children were trapped in forced labour situations worldwide in 2021. \cite{ILO2022}.

In response to these unintended consequences of globalization, some countries are introducing new regulations in the form of supply chain due diligence regulations. \cite{smit2020study, weihrauch2022voluntary}. The rationale for many of these rules is that voluntary agreements failed and international agreements cannot be enforced abroad by foreign governments. Therefore, companies that operate internationally are being held accountable. This means that the monitoring and enforcement of public regulations is delegated to private companies. Affected companies will need to adapt and update their compliance, sourcing and contracting processes. As risks can change, obligations must be monitored and improved on a regular basis. Both France, with the Loi de Vigilance in force since 2017, and Germany, with the Lieferkettensorgfaltspflichtengesetz in force since 2023, have passed such legislation.

To avoid a fragmentation of due diligence requirements across the EU Single Market, the EU proposed in 2022 a Directive on Corporate Sustainability Due Diligence (CSDDD; 2022/0051/COD))). The proposed directive aims to improve corporate governance practices to mitigate adverse HR and environmental impacts, to remedy adverse impacts for those affected and to promote sustainable and responsible business conduct throughout the global value chain. Firms operating in the EU need to ensure that they abide to high ethical, environmental and labour standards throughout their operations. The EU proposal of the CSDDD applies to all large companies operating in the EU, i.e. firms with more than 500 employees and an annual yearly turnover of more than €150 million world wide\footnote{Note that these thresholds are still discussed in the EU Trilogue. While the EU Council favors a higher threshold in terms of number of employees and turnover, i.e. less directly affected companies, the EU Parliament favors lower thresholds, with firms affected if they have more than 250 employees or an annual worldwide turnover of more than €150 million.} In addition, non-EU firms that generate more than €150 million of their annual net sales in the EU will also have to comply. There is a special focus on companies operating in "high impact sectors". These are sectors identified by the EU as having a high risk of negative impacts on the EU and a high potential for violations of human rights and environmental standards. They include wholesale trade of textiles, clothing and footwear, wholesales trade of agricultural raw materials, live animals, wood, food and beverages, agriculture, forestry, fisheries, extraction of mineral resources, manufacturing of food products and beverages, manufacturing of textiles, leather and related products and manufacturing of basic metal products, other non-metallic mineral products and fabricated metal products.   

The adoption and implementation of due diligence in accordance with the proposed EU CSDDD is fraught with barriers and challenges, particularly with respect to the requirements to identify, mitigate and prevent HR abuses and adverse environmental impacts throughout the value chain. Although an increasing engagement of companies in international business has led to a larger number of firms reporting on corporate responsibility and supply chain due diligence to meet the demands of suppliers, buyers, investors, customers and regulators, many companies do not yet comply with the required due diligence practices as proposed by the EU CSDDD \cite{meyerDoingWellDoing2022}. The cost of implementing and monitoring CSDDD depends on the level of the supply chain, the industry and the location of trading partners. 

Firms have to ensure that their direct and indirect suppliers comply with due diligence requirements, which can be administratively burdensome and even bureaucratic for many companies. Even when this supply chain risk management and monitoring is effectively implemented, many firms do not have a complete picture of their entire supply chain, leading to the possibility of inefficient mitigation, prevention, and exclusion of human rights abuses from their supply chain. 

Against this background, this study aims to provide an ex-ante assessment of the impact of supply chain due diligence regulations such as the proposed EU CSDD Directive. Assessing the effects of such regulations is difficult due to the lack of available data on international firm-level production networks. Therefore, we construct a full-scale network model of firms that screens the supply network for HR violations and provides the likelihood of a firm violating the regulations. 

The starting point is information about domestic firm networks. Recently, there has been an increase in the availability of country-wide data on firm-level production networks, which has been used for systematic analysis. For instance, several studies have made use of administrative VAT data~\cite{diemQuantifyingFirmlevelEconomic2022a,mungoReconstructingProductionNetworks2023,grigoliIdiosyncraticShocksAggregate2023,diem2023estimating}, or information from payment systems~\cite{silvaModelingSupplyChainNetworks2020,fujiwaraMoneyFlowNetwork2021,ialongoReconstructingFirmlevelInteractions2022}, or other sources such as telecommunications data ~\cite{reisch2022monitoring}, providing a comprehensive coverage of virtually all companies and their relationships within a national economy.
These studies uncover topological features of production networks at the firm-level~\cite{BacilieriFirm}.
For example, firms typically have an average degree of between 30 and 50 links with other firms.
This number scales with an exponent of approximately 1/3 with firm size.
Both, the out-degree and in-degree distributions exhibit an approximate power-law with tail exponents of around 1.5 and 2.5, respectively. This indicates that large firms generally have many more supply links than small ones.\footnote{Also the volume of monetary flows along these links show power-law statistics.}
The consistency of these and other findings across countries has been established by comparing data from Ecuador, Hungary, Belgium, Brazil, and Japan with data gathered from financial reporting obligations, shipment information, import-export data, and VAT data.

Another network property that is  consistent across countries is the characteristic path length, defined as the average length of the shortest path (network distance) between any two firms in the production network~\cite{BacilieriFirm}.
Considering networks with complete visibility of all links, the characteristic path length is found to be in the range between three and four. These values coincide with the expected values for the so-called configuration model, i.e., random networks with the same number of nodes, links, and degree sequence (same degree distributions).

In this study, we demonstrate that the topological characteristics of firm-level production networks have significant and far-reaching implications for supply chain due diligence in Europe. To illustrate these effects, we develop a detailed random network model based on EU-wide structural business statistics, input-output (IO) data, and export information. The model exhibits all pertinent topological properties (firm size, degree, and strength distributions, as well as their interdependencies) observed in the data and ensures consistency with inter-industry relationships within the European Union, as derived from IO data, and with imports from non-European countries, as derived from international trade data.

We link this model with two databases regarding HR violations in non-EU nations.
The first database is a list of goods produced by forced or child labour, supplied by the United States Department of Labor \cite{ListofGoods}.
The second is a HR allegation lawsuits database maintained by the Business and Human Rights Resource Centre \cite{HRLawsuits}.
Within our framework, we can assess the probability of an EU company having a direct or indirect connection with a non-EU company located in regions and sectors globally recognized for HR violations.
For various nations, industries, and business sizes, we measure the risk using a novel Supply Chain Due Diligence (SCDD) Risk Indicator. This indicator calculates the probability of locating a potential HR violator within a set distance in the production network of the firm.
Additionally, this model is utilised to create profiles for businesses of a particular size, sector and nation, that define the areas of origin most prone to potential links to human rights breaches. This is quantified by employing a new SCDD Exposure Indicator.

In brief, these indicators can be computed as follows.
Let $A$ be the unweighted directed $N\times N$ adjacency matrix of the full supply network of $N$ firms and $v$ be a vector indicating whether company $i$ is a potential HR violator, $v_i=1$ or not, $v_i=0$. Let $A_{ij}$ encode whether company $j$ supplies $i$, $A_{ij}=1$, or not, $A_{ij}=1$.
For each company $i$ we can then define the SCDD Risk Indicator for having a HR violator at tier $k$ in the supply chain as (in vector notation)
\begin{equation}
\text{SCDD Risk}(k)_i= \left(A^k v \right)_i > 0 \quad,
\label{SCDDRiskIntro}
\end{equation}
with $\text{sgn}$ indicating the component-wise sign function.
To obtain the indicators for a set of companies (e.g., a country, sector, or size bracket), we compute the average value of the company indicator in Eq.~\ref{SCDDRiskIntro} taken over all companies in the corresponding set.

\section{Results}
The reconstructed firm-level production network contains $N=30\,785\,000$ nodes with a standard deviation (SD) of $49\,00$ and $L=893\,900\,000$ (SD $7\,500\,000$) links, measured over multiple realizations. 

\begin{figure}[htbp!]
    \centering
    \includegraphics[width=0.49\textwidth]{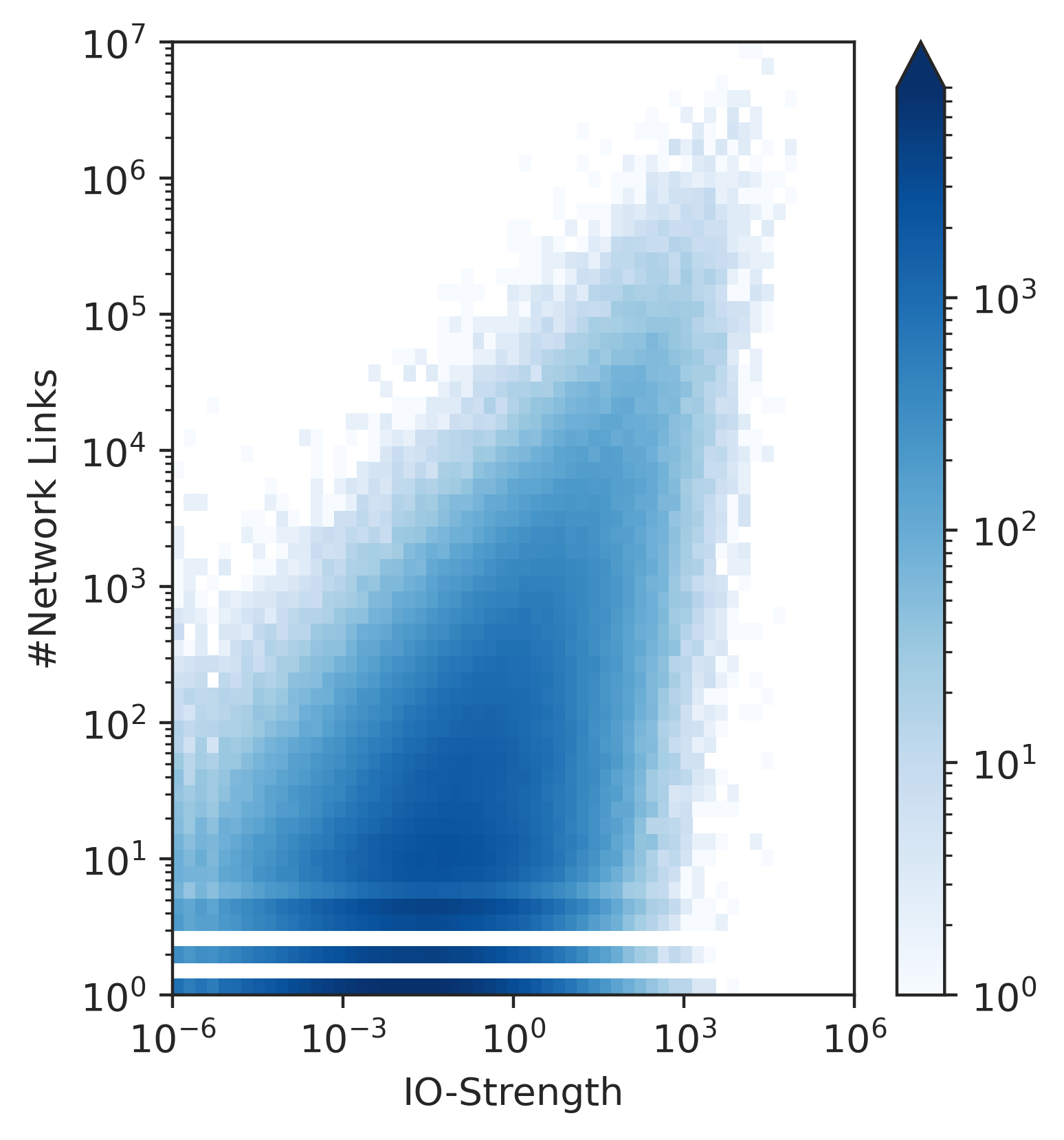}
    \includegraphics[width=0.49\textwidth]{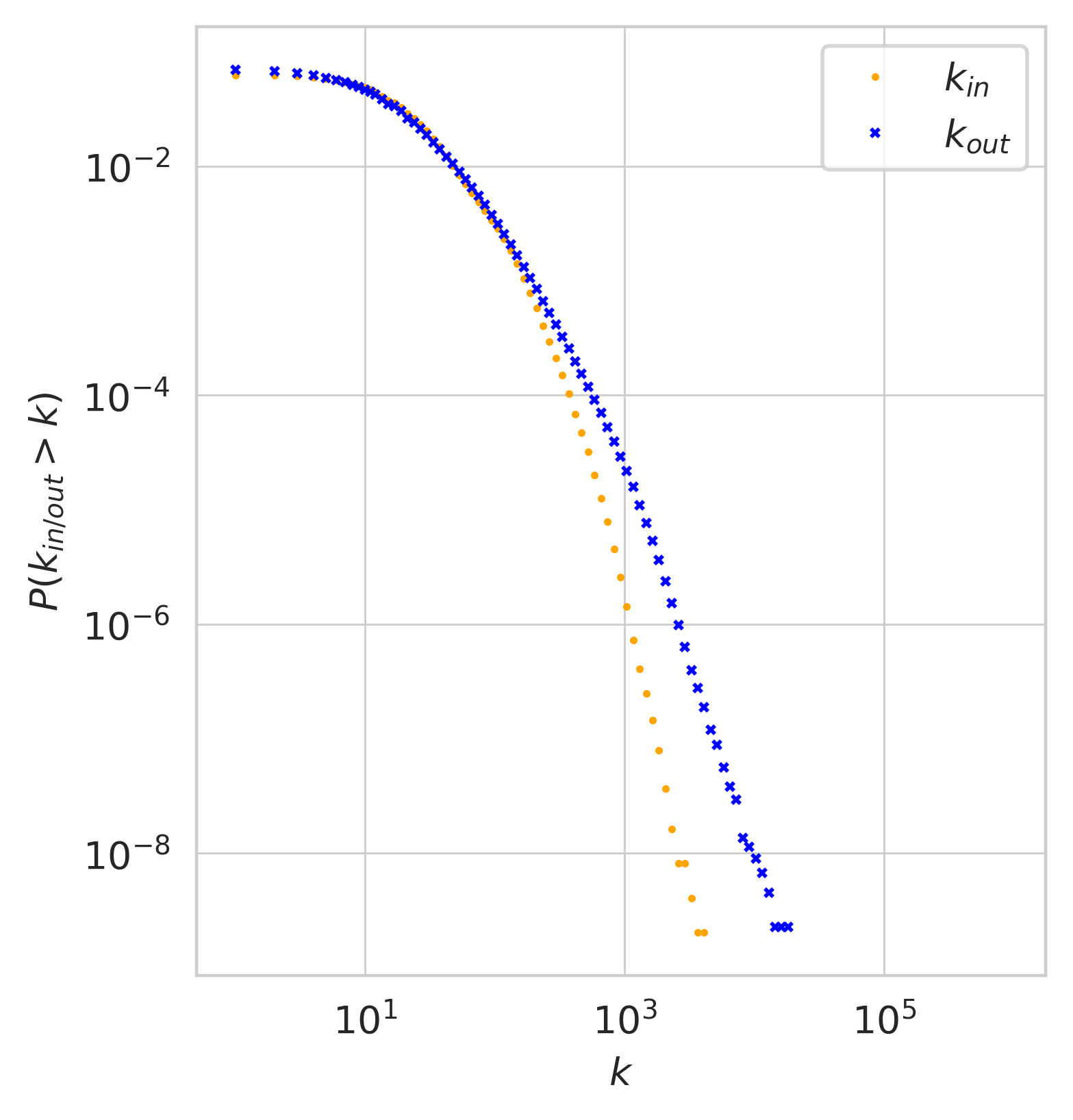}
    \caption{The strength between the (sector,country) pairs in the OECD IO-table against the number of links in the reconstructed networks (left) and the cumulative probability density function of the in-/out-degree of the reconstructed network (right).}
    \label{fig:network_validation}
\end{figure}

To show that the reconstructed networks have similar statistical characteristics as the real firm-level production network, we compare the inter-industry relationships between the multiregional IO data and the network in Figure~\ref{fig:network_validation}. We show the value of the relations between each (sector,country) pair in the OECD IO table against the number of links in the reconstructed network, showing a positive correlation. The substantial variation in this plot (note the logarithmic axes) stems from the heavily right-skewed distribution of company sizes, which is also apparent in the in- and out-degree distribution of the reconstructed network. Both of these distributions show properties of fat tails as expected from the data ~\cite{BacilieriFirm}, with the out-degree distribution (number of customers per company) showing a more pronounced tail than the in-degree distribution (number of suppliers per company).
In particular, large companies are expected to have multiple thousands of suppliers and multiple ten thousands of customers.

\begin{figure}[htbp!]
    \centering
    \includegraphics[width=0.95\textwidth]{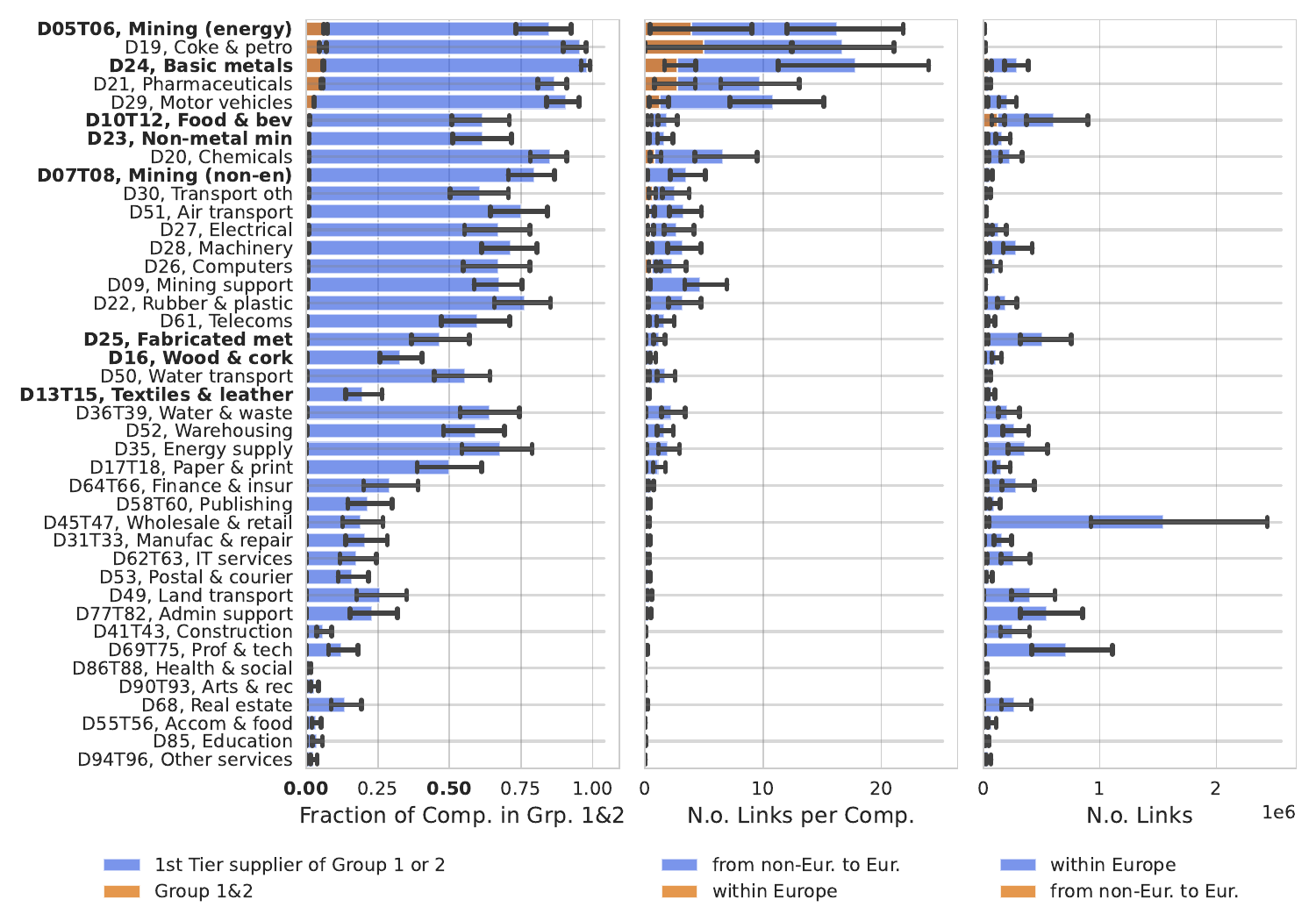}
    \caption{Estimation of firms and their direct suppliers for the EU CSDD Directive. For each sector (left panel), we show the fraction companies that would be required to monitor their supply chain under the currently proposed regulation (companies belonging to group 1 and 2, orange) as well as the fraction of companies that are estimated to be direct suppliers of those (blue). High risk sectors are highlighted in bold. We also show the number of links to monitor per company (middle) and in total (right panel).}
    \label{fig:group12}
\end{figure}

The EU CSDD Directive, as it is currently proposed, should apply to all EU companies of substantial size and economic power (with more than 500 employees and EUR 150 million in net turnover globally, group 1) or  companies operating in defined high impact sectors with more than 250 employees and a net turnover of EUR 40 million worldwide (group 2; for these companies, rules will start to apply 2 years later than for group 1). We estimate that only a small fraction of companies (about 20,000) belongs to these two groups, see Figure\ref{fig:group12}.
For many sectors, however, most companies are direct (tier 1) suppliers of a group 1 or 2 company, suggesting that the number of companies indirectly affected by the EU CSDD Directive is orders of magnitude larger than the companies to which it applies.
In some sectors, these companies have up to 20 links that they need to monitor, which amounts to up to a million of links per sector that require monitoring.

\begin{figure}
    \centering
    \includegraphics[width=0.95\textwidth]{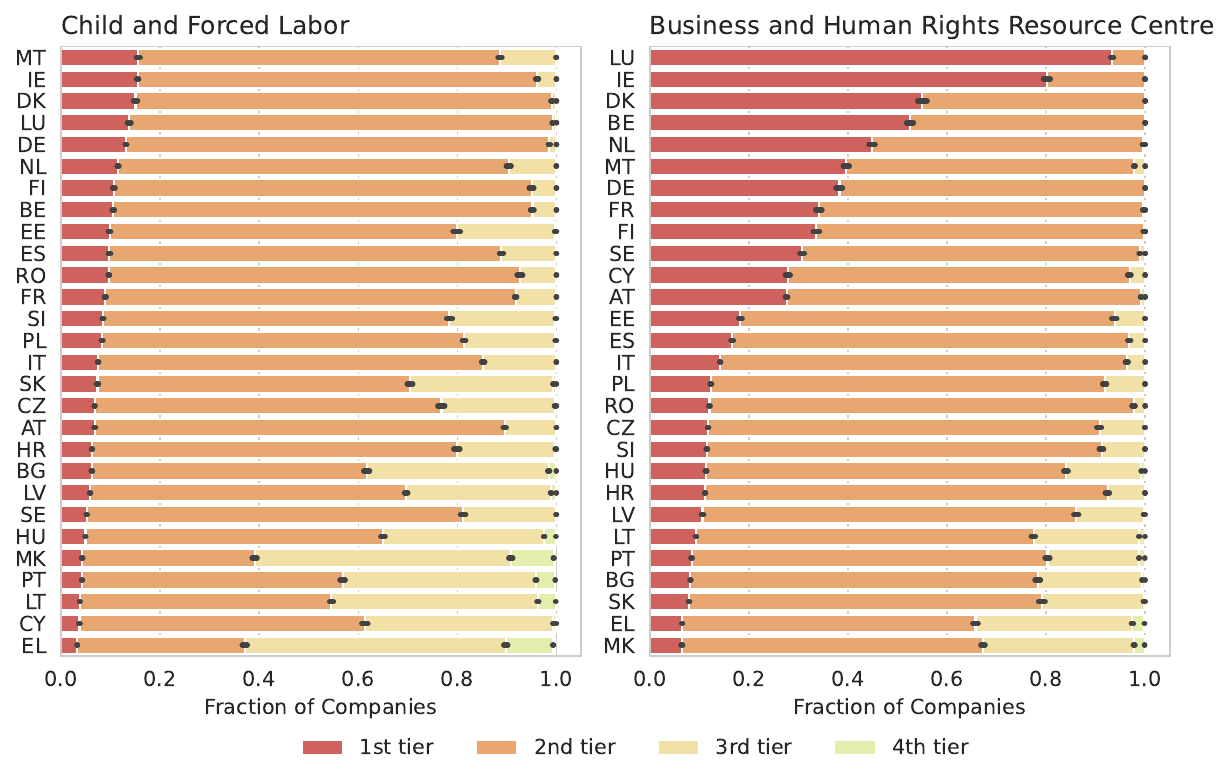}
    \caption{Results for the SCDD Risk Indicator on a country level. Results are shown for indicators on tier one to four (colors) for (a) allegations of child an forced labor as well as (b) lawsuits collected by the Business and Human Rights Resource Centre. Tier one indicators are lower for the child and forced labor dataset compared to the lawsuits. Countries including Luxembourg and Ireland show the highest tier one risks. Considering the third or higher tiers, all indicators are close to 100\%.}
    \label{fig:country_results}
\end{figure}

We now consider the likelihood of having a potential HR violator not as a direct supplier, but on a higher tier in the supply chain. Figure~\ref{fig:country_results} presents the results of the SCDD Risk Indicator at the country level.
Risk is measured across four tiers of the supply chain, which correspond to the number of steps one must take through the directed firm-level production network to reach a non-EU firm in a sector posing a high risk of HR violations. 
Results are shown for (left) child and forced labour allegations and (right) lawsuits recorded by the Business and Human Rights Resource Centre. 
We observe significant disparities among countries on the first tier (referring to direct neighbours in the network).
Some countries with the highest tier one risks are very similar in both datasets, namely Luxembourg, Ireland, Malta, Denmark, Belgium, Netherlands or Germany.
Regarding child and forced labour, we find that about 10\% of companies in these countries have a direct association with a problematic sector, whilst for most other countries this risk remains below 10\%.
The tier one indicators are higher for the lawsuits database compared to the child and forced labor dataset, with more than 60\% for Luxembourg and Ireland.
The second tier SCDD Risk indicator is around 40 and 60\% or higher in all EU nations and the two datasets, respectively.
For tiers of three or higher, the SCDD Risk Indicator reaches almost 100\% for all businesses EU-wide.

\begin{figure}[htbp!]
    \centering
    \includegraphics[width=0.95\textwidth]{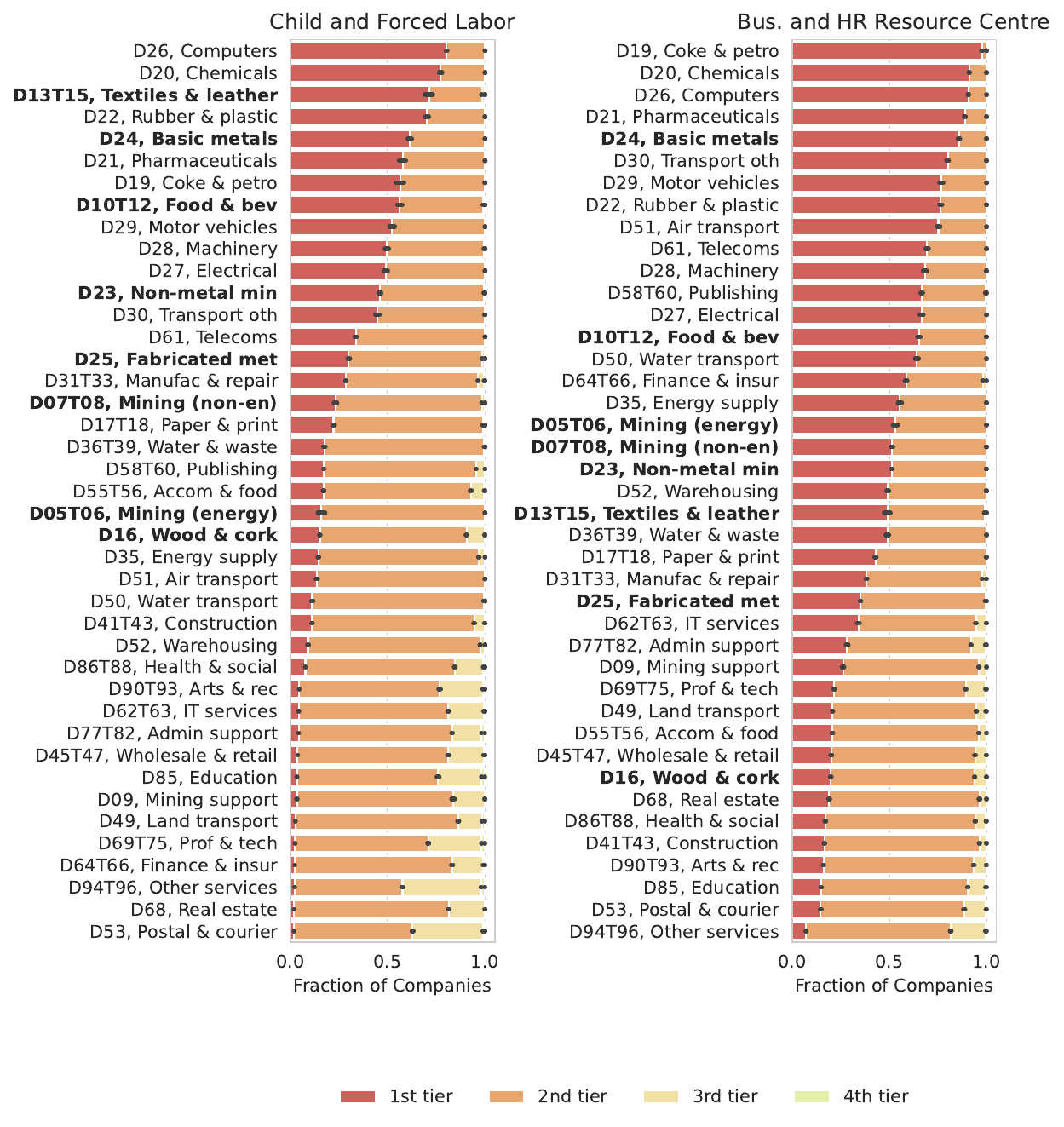}
    \caption{Results for the SCDD Risk Indicator on a sector level. Results are shown for indicators on tier one to four (colors) for (a) allegations of child an forced labor and (b) lawsuits. Rankings of sectors by tier one indicators are similar across both datasets, with highest values for computers, basic metals, chemicals, pharmaceuticals, motor vehicles and textiles. Considering the 4th or higher tiers, all indicators are close to 100\%.}
    \label{fig:sector_results}
\end{figure}

Figure~\ref{fig:sector_results} displays the results of the SCDD Risk Indicator at the sector level for (left) forced and child labour and (right) lawsuits.
As with the country-level results, all indicator values are almost 100\% for tier three and above.
However, an examination of the tier one ranking reveals notable differences between specific sectors.
High-risk sectors identified by our analysis comprise of coke and petroleum, basic metals, textiles and leather, as well as the manufacturing of computers, chemicals and pharmaceuticals. These identified sectors are related directly or through intensive IO linkages to the high-impact sectors as defined by the EU CSDDD proposal, see Figure~\ref{fig:sector_results}. 
On the other end of the spectrum, we mostly find service-oriented sectors like arts, recreation, postal services, health and social services, real estate, and research.
These results prove to be fairly robust across two independent datasets.

Companies that employ over 50 people exhibit substantially more elevated indicator values in comparison to smaller enterprises. 
Concerning larger firms with more than 250 employees, the tier one risks are mainly within the ranges of 30\% to 50\% for child and forced labour and 40\% to 90\% for the lawsuit dataset, see Figure~\ref{fig:size_geo} and \ref{fig:size_sec} in the Supplementary Material.
When examining the tier one indicator results based on firm size and sector, it is evident that risk increases as firm size increases. 
There are significant differences between sectors, compared to the country-level outcomes.
In the case of large companies, there are numerous high risk sectors, namely computers, basic metals, and chemicals, and so forth, which all demonstrate indicator values of over 60\% or 80\% for child and forced labour allegations or HR lawsuits, respectively.


\begin{figure}
    \centering
    \includegraphics[width=0.95\textwidth]{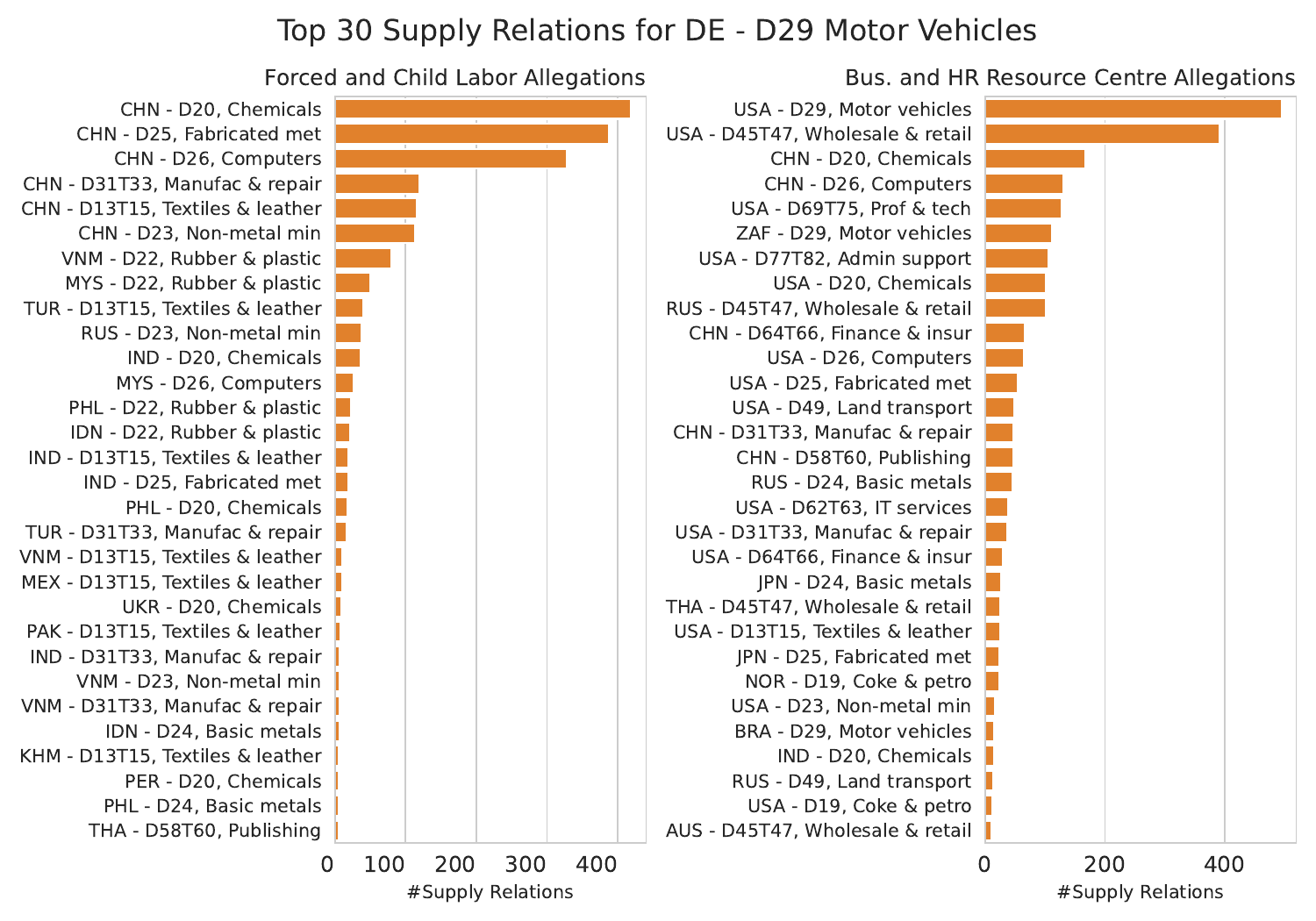}
    \caption{The SCDD Exposure Indicator for the manufacture of motor vehicles in Germany (DE - D29) with child and forced labour allegations and lawsuits recorded by the Business and Human Rights Resource Centre. Companies in this sector have the highest exposures to chemicals, fabricated metals, textiles and leather from China.}
    \label{fig:de_c29}
\end{figure}

Within our framework, the SCDD Exposure Indicators can be computed for all combinations of sectors, countries, and firm size categories.
While a full exploration of these indicators is outside the scope of this article, we will provide detailed results for selected sectors.

In particular, we examine the production of motor vehicles in Germany as a case study.
Figure~\ref{fig:de_c29} illustrates the results of the SCDD Exposure Indicator for this sector concerning child and forced labour.
It is apparent that the SCDD exposure is concentrated on several other sectors, notably fabricated metal, textiles and leather, machinery manufacturing and repair, computers, chemicals, and non-metallic minerals in China.
There is also a slightly higher exposure associated with rubber and plastics from Vietnam and Malaysia as well as textiles and leather products originating from Turkey.

However, a very different picture emerges from the results for litigation, see Figure\ref{fig:de_c29}.
Specifically, the automotive and wholesale and retail sectors in the USA show the highest concentration of exposure for Germany, followed by chemicals and computers from China.
Note that these associations stem from specific lawsuits.
Firstly, there is an ongoing class action lawsuit against Walmart, in which women from across the US have alleged gender-based discrimination \cite{walmart_ls}.
The link to the wholesale and retail sector in Russia further concerns the "Golyanova slaves" \cite{golya_ls}, in which former employees of the grocery chain claim they were forced to work in slave-like conditions, subjected to forced labour, confiscation of their passports and physical abuse. 
Lawyers filed a complaint with the European Court of Human Rights after the local court refused to hear the case. 

\begin{figure}[htbp!]
    \centering
    \includegraphics[width=0.8\linewidth]{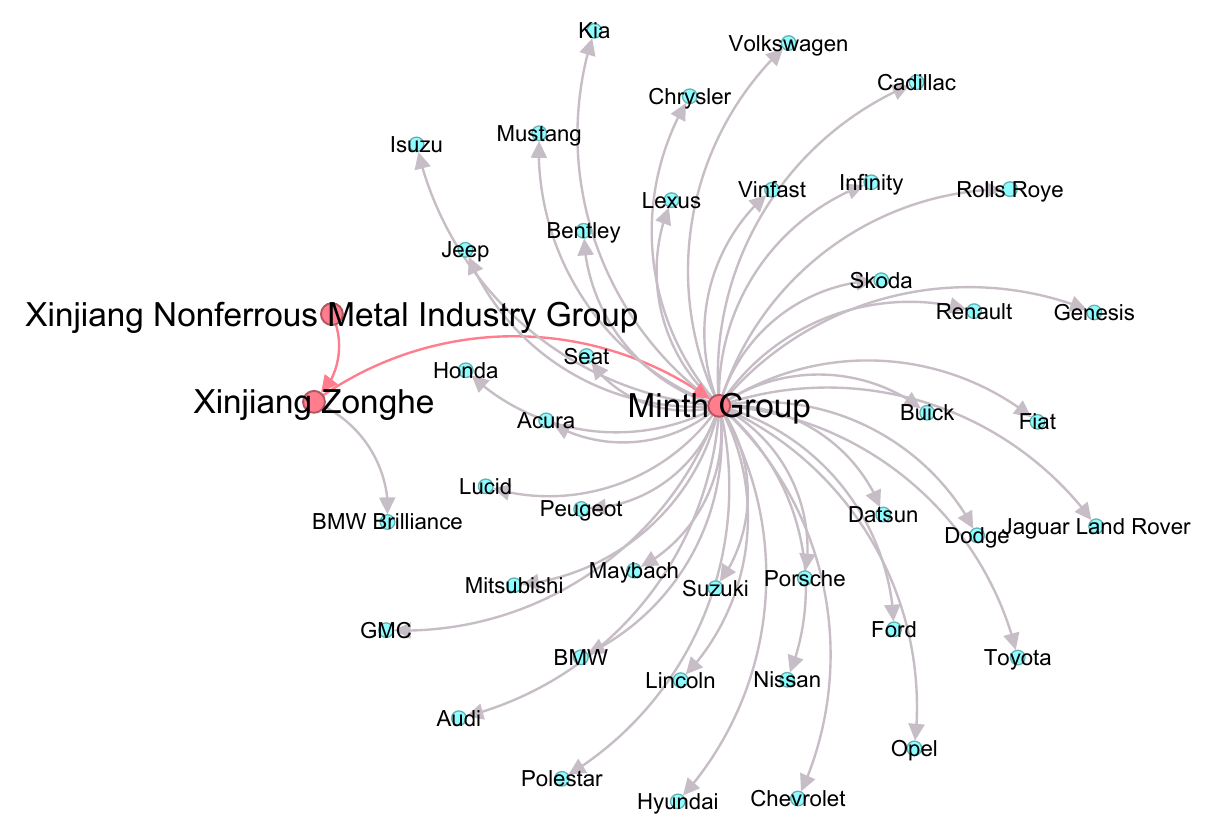}
    \caption{Real-life example of the small world effect in supply chain due diligence. Xinjiang Nonferrous Metal Industry Group (red) is a company located in the Uyghur autonomous region linked to forced labour. It provides critical raw materials for many companies in the automotive industry or other suppliers of companies therein, shown as turquoise nodes. Most car manufacturers can be linked to the Xinjiang Group by supplier relationships (links) in no more than three steps on this network.}
    \label{fig:uyghur}
\end{figure}

To illustrate our main result, that the question is not whether a firm has an HR violator in its supply chain or not, but rather how many violators there are and at what level, we take a deeper look in real-life - rather than stylized - supply chains. 
Xinjiang Nonferrous Metal Industry Group is a state-owned mining, smelting, and processing company in the Uyghur autonomous region supplying some of the automotive industry’s most critical raw materials, including copper, zinc, lithium, gold, and nickel. Via one of its subsidiaries, Xinxin Mining, it is a supplier to Xinjiang Zhonghe Co., Ltd. Xinjiang Zonghe is an aluminum smelter, alloying company and products manufacturer producing 180,000 tons of aluminum annually. This makes Xinjiang Zonghe the world’s largest manufacturer of high-purity aluminum, exporting products to Japan, Europe, South Korea and the US. Xinjiang Zonghe is a direct supplier of BMW Brilliance (BMW’s joint venture with China Automotive Holdings). Additionally, Xinjiang Zonghe is a direct supplier of Minth Group. Minth Group designs and manufactures structural bodies, trims and decorative parts for the automotive industry. Minth operates over 50 production plants supplying automotive markets in 30 different countries. Minth is a direct supplier to almost all OEMs worldwide (see Figure~\ref{fig:uyghur}). Xinjiang Nonferrous Metal Industry Group, Xinjiang Xinxin Mining Industry Co., Xinjiang Zonghe as well as the Minth Group are directly linked to Uyghur forced labor \cite{JWW}.
Hence, most car manufacturers around the globe can be linked to Uyghur forced labor in no more than three tiers in their supply chain, according to this analysis.

\section{Discussion}

In this work, we study the impact of the small world effect in firm-level production networks for supply chain due diligence.
We present two indicator sets to measure the risk and exposure in supply chain due diligence for EU firms.
We reconstructed the firm-level EU production network using a random network model that is in full compliance with EUROSTAT structural business statistics, multi--regional IO data, export data, and various stylized statistical facts that have been consistently established across several actual nation-scale company-level production networks.

We propose the SCDD Risk Indicator to quantify the probability of a link to a potential HR violator at a given tier in a company's supply chain. This is measured through comprehensive databases of sectors with high risks of child and forced labor and HR violations.
Additionally, the SCDD Exposure Indicator quantifies the most pertinent sources of this risk for a given group of companies.
Using these indicators, we show that almost all European companies are no more than three steps away from the potential HR violator in the international supply network.

Our findings have significant implications for the structure of effective regulations for supply chain due diligence. In particular, it is necessary to acknowledge that firms' global production networks are dense and clustered, leading to a well-established network phenomenon referred to as "small worlds".
It has been observed that individuals within social networks are separated by no more than six degrees of separation from any other individual on the planet, commonly known as the small world effect \cite{watts2004six}.
This phenomenon has been validated in several social networks, including Stanley Milgram's pioneering letter-passing experiment \cite{milgram1967small} and studies of massive online communication networks \cite{dodds2003experimental,leskovec2008planetary}.
We show that on a European scale firm-level production networks can be classified as small worlds.

Previous research established that firm-level production networks have an average network degree in the range of 30 to 50 in addition to a tendency to form hubs, i.e., nodes with a degree that is orders of magnitudes higher than the average degree.
These hubs often correspond to large multinational companies that are known to have tens of thousands of direct suppliers.
In many cases, they apply a dual or multiple sourcing strategy which further multiplies their number of direct suppliers.

The emerging small world effect in such networks has immediate consequences for supply chain due diligence. Even if there were only one HR violator in the entire network, it is still possible that a significant part of the world economy has this violator at some tier within its supply chain.
Our research reveals that production networks are a smaller world than social networks, with no more than three degrees of separation between each European company and potential non-EU HR violators.
Expanding a firm's responsibility from its internal production processes to its entire supply chain immediately expands this responsibility to the worldwide production network.

Mining and quarrying, along with the production of apparel, are considered as "high impact sectors", i.e., as  sectors with the highest risks for supply chain due diligence issues. The manufacture of basic metals and agriculture also pose significant challenges.
This is supported by the fact that these sectors have stricter reporting thresholds for due diligence assessments, in terms of lower threshold number of employees and revenue, in the proposed EU legislation \cite{scdd_regulation}.
While our findings support the substantial risk linked with these sectors, they also identify other areas having equally high SCDD Risk Indicators. Specifically, the production of computers, chemicals, pharmaceuticals, and motor vehicles exhibit comparably elevated tier one indicator values. These are sectors highly interlinked with the high-impact sectors. 
Therefore, our model suggests that identifying only mining, minerals, basic metals and textiles as high-risk sectors is untenable from a data perspective.

Taken together, our results suggest that effective supply chain due diligence regulation should consider the intricacy of the network phenomena that are inherent in firm-level production networks.\cite{BacilieriFirm}.
The currently proposed EU regulation implies that due diligence efforts should be focused on the links in the networks and that the nodes (firms) are responsible for this assessment. A firm would be liable for any node along its supply chain. This implies both far-reaching liabilities and substantial surveillance costs for any internationally active firm, providing a lever for policy adjustment.
A straightforward simplification of the regulation is to reduce the liability of firms to the first tier. 
Firms would only be responsible for the conduct of their first-tier suppliers, unless the companies have knowledge of downstream SCDD violations. 
According to our estimates, however, this would still require to monitor practically all companies and million of supply links in some sectors.

Alternatively, due diligence efforts could focus only on the largest links.
This is problematic for several reasons.
First, it would require setting a threshold below which it would be considered acceptable to maintain business relationships with HR violators.
Second, due to the high density of production networks at the firm level, this threshold would have to be set at rather high values to avoid the emergence of the small-world effect described above.
More specifically, in network theory terms, the crucial property is the emergence of a so-called strongly connected component (SCC), which is defined as a set of nodes in a directed network for which each node can be reached from any other node in the component by following a path on the network.
For random networks with homogeneous degree distributions, a SCC can be expected if there is at least one incoming and outgoing link for each node, implying an average degree of one for both the incoming and outgoing degree distributions.
Thus, with an average degree greater than 30, the thresholds for the application of the EU CSDDD would have to be lowered to the point where the vast majority of links are excluded from the assessment if the small-world effect is to be avoided, thereby diluting the policy.
Note also that firm-level networks were found to exhibit a remarkable degree of temporal stability \cite{borsos2020unfolding}, with about 50\% of links persisting for more than one year. 

An alternative approach to supply chain due diligence would be to designate authorities to maintain registers of compliant and non-compliant companies. In this case, instead of carrying out assessments for each link in the network, assessments would be carried out for each node, reducing the number of assessments by a factor equal to the mean degree in the network. To further reduce the complexity of the network to be monitored, geographical areas with sufficient labour standards such as the EU itself could be exempted from the regulation (positive lists of companies).
Companies that have been identified as repeat offenders could be placed on negative lists and thereby explicitly prohibited from participating in EU supply chains.
The EU CSDD Directive could then only be applied in cases where companies are neither on a positive list nor on a negative list.

The benefits of node-based rather than link-based due diligence monitoring can be estimated quantitatively.
For example, based on our model, we estimate that there are 20,000 EU-based companies under the EU CSDDD. 
These companies are estimated to have 5 million suppliers along 9 million supply relationships. Further, 30 million EU companies and virtually all global companies importing into the EU are no more than three levels in the supply network away from the original 20,000 companies, giving a maximum of 900 million supply relationships that could need to be monitored under the CSDDD without adopting a threshold below which relationships are exempt from monitoring.
Thus, node-based rather than link-based monitoring would reduce the total number of required due diligence assessments by a factor of 30 without compromising the Directive.
According to our model, we further estimate that there are approximately 750,000 supply links from non-European entities to entities covered by the CSDD Directive. 
Focusing efforts on these companies (assuming EU companies are placed on a positive list) could again reduce the total number of required assessments by several orders of magnitude.

A more centralized, policy-driven approach to monitoring due diligence in supply chains would be to link retrievable information (i.e., financial flows between companies) with publicly available information about companies that violate agreed codes of conduct. This saves financial and human resources at the corporate level and eliminates the need to build overlapping production networks at the firm level. Without revealing sensitive business information, government agencies can conduct due diligence and proactively inform companies that may be at risk of misconduct in their supply chain.

Our study has certain limitations.
First, the outcomes are based on a probabilistic network model that was calibrated to sector-level information. Statements concerning the due diligence of individual firms in their supply chains cannot be made, as data on the actual production network at the company level throughout Europe would be necessary to do so.
Furthermore, our model might differ from actual situations if there are structural features that deviate significantly from our modeling assumptions.
However, recent research on countrywide networks between firms has demonstrated that the configuration model and variations that employ node strengths instead of node degrees, generally provide a realistic generative model \cite{ialongoReconstructingFirmlevelInteractions2022}.
Furthermore, we considered multiple samples of our model and discovered that the outcomes were remarkably robust across these samples. 
From this, we can infer that the outcomes of our indicators at the sector and country levels are genuinely reflective of the real-world network. However, no conclusions can be made regarding individual businesses; our results are of statistical nature.

Second, our indicators measure whether EU companies are linked to any company in a "problematic" non-EU sector.
It is important to note that this modelling choice was made due to data limitations, as reliable estimates for the fraction of companies within these sectors associated with HR violations are not available.
Nonetheless, the small world effect operates very likely also within these non-EU industries. This implies that even if a small number of violators exist within a specific sector, it is likely that a large majority of other firms are also directly or indirectly related to these offenders.

Third, it is important to acknowledge that limitations exist regarding the completeness and accuracy of the two HR violation datasets, which may also extend to our method.
Over- or underrepresentation of certain sectors or countries in these datasets could lead to a distortion of our indicator values.

In summary, our findings from a synthetically generated EU supply network challenge some of the key assumptions underpinning the current proposed regulation on supply chain due diligence within the EU.
Supply chains of individual companies cannot be isolated from the rest of the network.
Therefore, instead of focusing efforts on individual companies that have to monitor all their direct and indirect links, as currently proposed in the EU CSDD Directive, our results show that the costs and burdens of due diligence regulations can be effectively pooled by focusing efforts only on companies, i.e. by monitoring the nodes rather than the links in the supply network.

\section{Methods}
\subsection{Data}
\subsubsection{Eurostat - Structural business statistics}
Data from Eurostats structural business statistics~\cite{EurostatBusiness}
includes information on companies in all EU-countries (and Iceland, Norway and Macedonia) sector and employees size groups.
Sectors are given on a NACE 2 digit level starting at B05, which means that agricultural sectors are not included.
The size groups are defined in the ranges of 0-9, 10-19, 20-50, 50-150, 150-250 and more than 250 employees.
20\% of the values are missing; these represent around 2\% of the total turnover and, however, are concentrated in specific sectors and countries, where they can represent more than 50\% of the sector turnover.
\subsubsection{OECD - IOTs}
Data on intersectoral relationships were drawn from the 2021 edition of the OECD Input-Output Tables~\cite{OecdIot}. These cover 67 countries each consisting of 44 industries, with 2018 being the most recent year covered. 
Notably, all EU countries are included, and they provide a mapping to the NACE level 2 sectors.


\subsubsection{Comtrade - Exports}\label{subsec: comtrade}
For non-European Union (EU) countries, we source all export data destined for EU member countries in the year 2021 from the United Nations Comtrade database, accessible at \cite{Comtrade}. Subsequently, we convert the Harmonised System (HS) codes first to International Standard Industrial Classification of All Economic Activities, Revision 3 (ISIC3), then to ISIC4 codes (and thereby two digit NACE codes) to map the trade data to economic sectors \cite{ConversionTable}.
This results in a table of the imports of each EU country by sector and country of origin.

\subsubsection{HR-violations Databases}
To estimate human rights allegations within non-European Union (EU) countries across various sectors, we employ a method that draws from two distinct databases:
First, we employ the "List of Goods Produced by either Child or Forced Labor" provided by the United States Department of Labor \cite{ListofGoods}.  To align these sectors with broader economic classifications, we perform a conversion of the HS codes to ISIC4 codes. 
In addition, we harness the database on human rights allegation lawsuits maintained by the Business and Human Rights Resource Centre \cite{HRLawsuits}. This database provides records of lawsuits pertaining to human rights violations across a range of sectors. To align these unique sectors with standardized economic classifications, we map these sectors to ISIC4 codes.

\subsection{Sampling}
The objective is to create a set of synthetic companies whose size distribution in terms of number of employees, turnover and in-/out-degree matches the data described above.
For every country $c$, sector $s$, and size group $b$, data on the number of companies, the average number of employees per company, and the average revenue per employee is available.
An assumption is made that the distribution of company sizes, measured by the number of employees per company, follows a heavy-tailed distribution \cite{BacilieriFirm, borsos2020unfolding}, which we parameterize as a Pareto Type II distribution. To optimize the parameters of shape and scale for this distribution, a grid search for each combination of $(c,s,b)$ is performed. During this search, we compute the expected average within the corresponding size group and select the parameter pair that results in an average number of employees closest to the observed number.
For each $(c,s,b)$ combination, we then generate a sample from the associated distribution. This sample provides us with a list of company sizes, quantified by the number of employees per company. We subsequently estimate the company's turnover from the estimated number of employees using the turnover per employee specific to the $(c,s,b)$ category.
To translate these into in- ($k_{in}$) and out-degrees ($k_{out}$), we use the following scaling relationships between node degrees and strengths found in~\cite{BacilieriFirm}:
\begin{align*}
    \log(k_{out}) &= \alpha_{out} \log(s_{out})  + \beta_{out}, \qquad  \alpha_{out} \approx [0.31,0.36] \\
    \log(k_{in}) &= \alpha_{in} \log(k_{out}) + \beta_{in}, \qquad \alpha_{in} \approx [0.6,0.8]
\end{align*} 
The out-strength $s_{out}$ is identified with the turnover.
To sample the out-degrees based on the turnover, we sample $\alpha^i_{out} \sim \text{Normal}(0.335,0.025)$ and calculate the out-degree $k^i_{out}$ of each company as $k^i_{out}= e^{\beta_{out}} \cdot (s^i_{out})^{\alpha_{out}}$ and similarly, for the $k_{in}$, we sample $\alpha^i_{in} \sim \text{Normal}(0.7,0.1)$ and consequently calculate $k^i_{in}= e^{\beta_{in}} \cdot  (k^i_{out})^{\alpha_{in}}$.
We set $e^{\beta_{in/out}}$ such that the average in-/out-degree is equal to $\bar{k}_{in} = 56$ and $\bar{k}_{out} = 50$~\cite{BacilieriFirm}. 
The in-degree is 13\% higher than the out-degree, to accommodate for the imports.

This procedure results in a list of approximaley 30 million companies, where each company is in a specific country and sector, has a specific size in terms of number of employees and turnover, and has a specific in- and out-degree $k^i_{in}$, $k^i_{out}$. 
To also include the imports from the rest of world into the EU, we append dummy variables for the rest of the world (ROW) to this list of companies.
These only have a specific sector and an out-degree of $k^i_{out} = 1$.
Each of these companies is meant to represent one in-link from outside of the EU to an EU-company.
The number of ROW dummies we add in each sector is given as $\sum_{i \in \text{EU}} k^{i}_{in} \cdot \allowbreak (\text{ROW inflow of the sector in the IOT})/\allowbreak (\text{Total inflow exclusive ROW of the sector in the IOT})$.
´
\subsection{Network reconstruction}
The network reconstruction algorithm aims to produce an unweighted directed network, where the in-/out-degree of each company most closely resembles the sampled $k^i_{in}/k^i_{out}$ above, and the wiring probabilities of companies in different sectors and countries is proportional to the intersectoral flows in the IO-table.
The employed algorithm is sketched below.
\begin{algorithm}[H]
\caption{Network reconstruction}
\begin{algorithmic}
\REQUIRE IOT,$k^i_{out}$,$k^i_{in}$
   \WHILE{ number of links per company < 30}
       \STATE According to IOT-strength, sample 2 sector country pairs: $(s,c)_{out}$,$(s,c)_{in}$
       \STATE From companies in $(s,c)_{out}$ sample one with probability according to their out-links $k^i_{out}$: $i_{out}$
       \STATE From companies in $(s,c)_{in}$ sample one with probability according to their in-links $k^i_{in}$: $j_{in}$
       \IF{link $i_{out} \to j_{in}$ not in Network}
           \STATE add  $i_{out} \to j_{in}$  to Network
           \STATE $k^{i_{out}}_{out} -= 1$ and $k^{j_{in}}_{in} -= 1$
       \ENDIF
      \ENDWHILE
   \RETURN network
\end{algorithmic}
\end{algorithm}
\subsection{EU Imports}
So far, the ROW dummies only have a sector of origin and, from the reconstructed network, we know to which EU country it links to. 
We then use the dataset derived from the Comtrade data \ref{subsec: comtrade} to assign each ROW dummy a country of origin, with a probability that is proportional to their total import values.

\subsection{SCDD Risk Indicator and SCDD Exposure Indicator}
The SCDD Risk Indicator is the fraction of companies per EU country, sector and/or size group which  have direct (1st tier) or indirect (2nd, 3rd and 4th tier) connections to companies which are a potential HR violator according to the two HR violations databases.

The SCDD exposure indicator tells us how many supply relations a sector in an EU country has with sectors in countries which are in the HR violations databases.

\FloatBarrier
\newpage
\bibliographystyle{ieeetr}  
\bibliography{manual_lit}

\newpage
\section{Supplementary Material}

\begin{figure}[h!]
    \centering
    \includegraphics[width=0.49\textwidth]{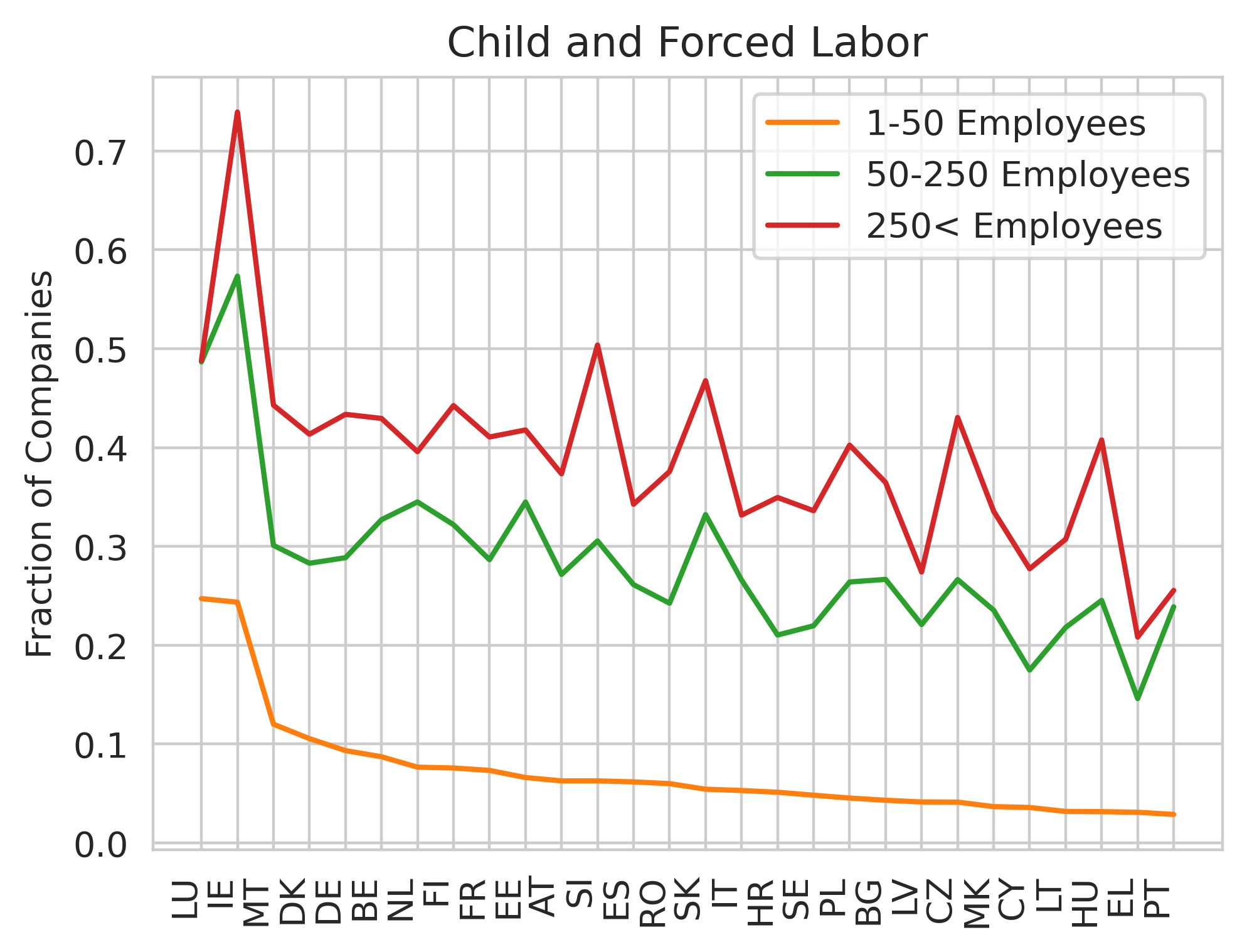}
    \includegraphics[width=0.49\textwidth]{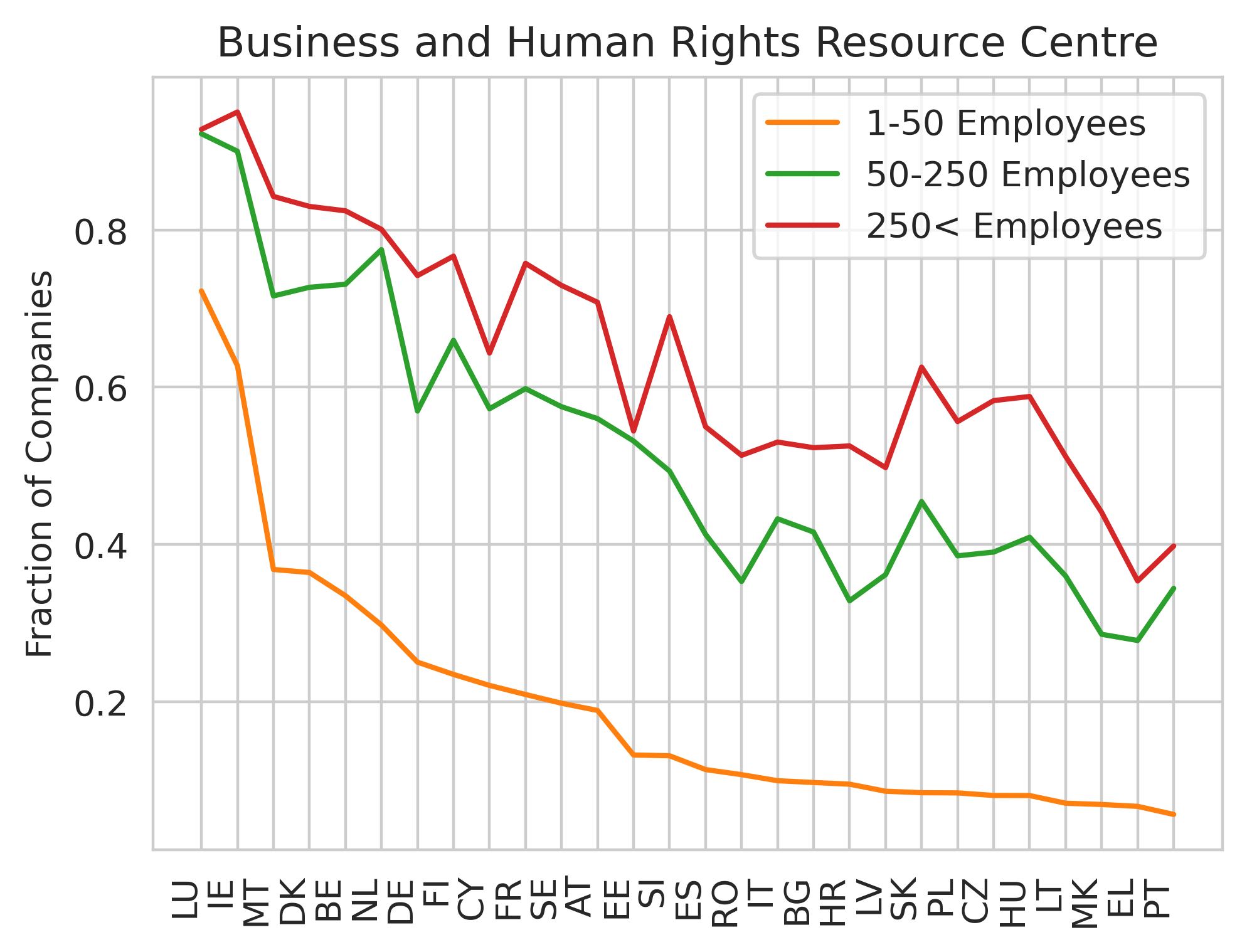}
    \caption{Results for the SCDD Risk Indicator by countries on tier one, stratified by firm size categories. Companies with 1--50 employees show substantially lower risks than larger companies for (a) child and forced labour allegations and (b) HR-related lawsuits.}
    \label{fig:size_geo}
\end{figure}

\begin{figure}[h!]
    \centering
    \includegraphics[width=1\textwidth]{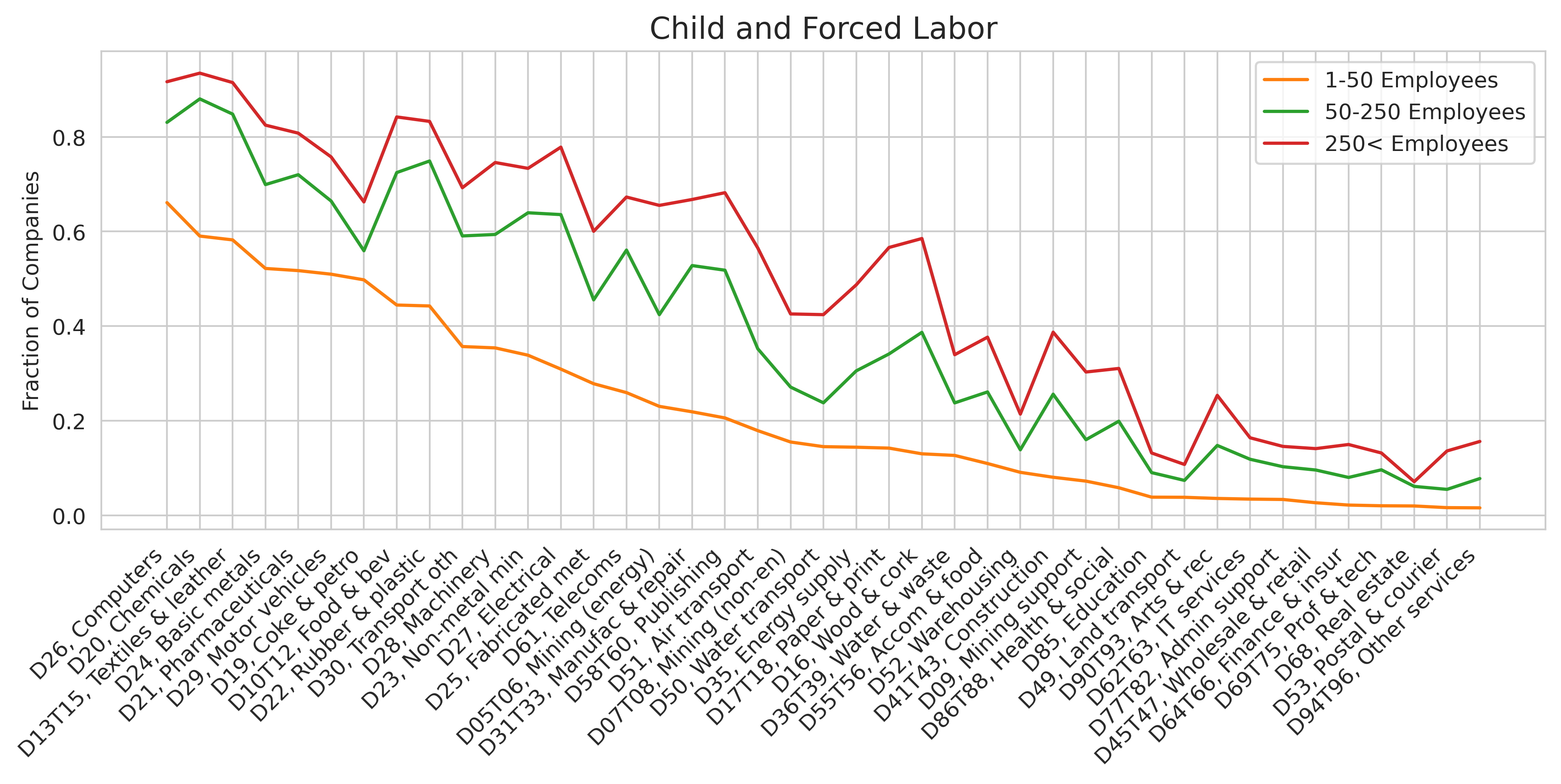}
    \includegraphics[width=1\textwidth]{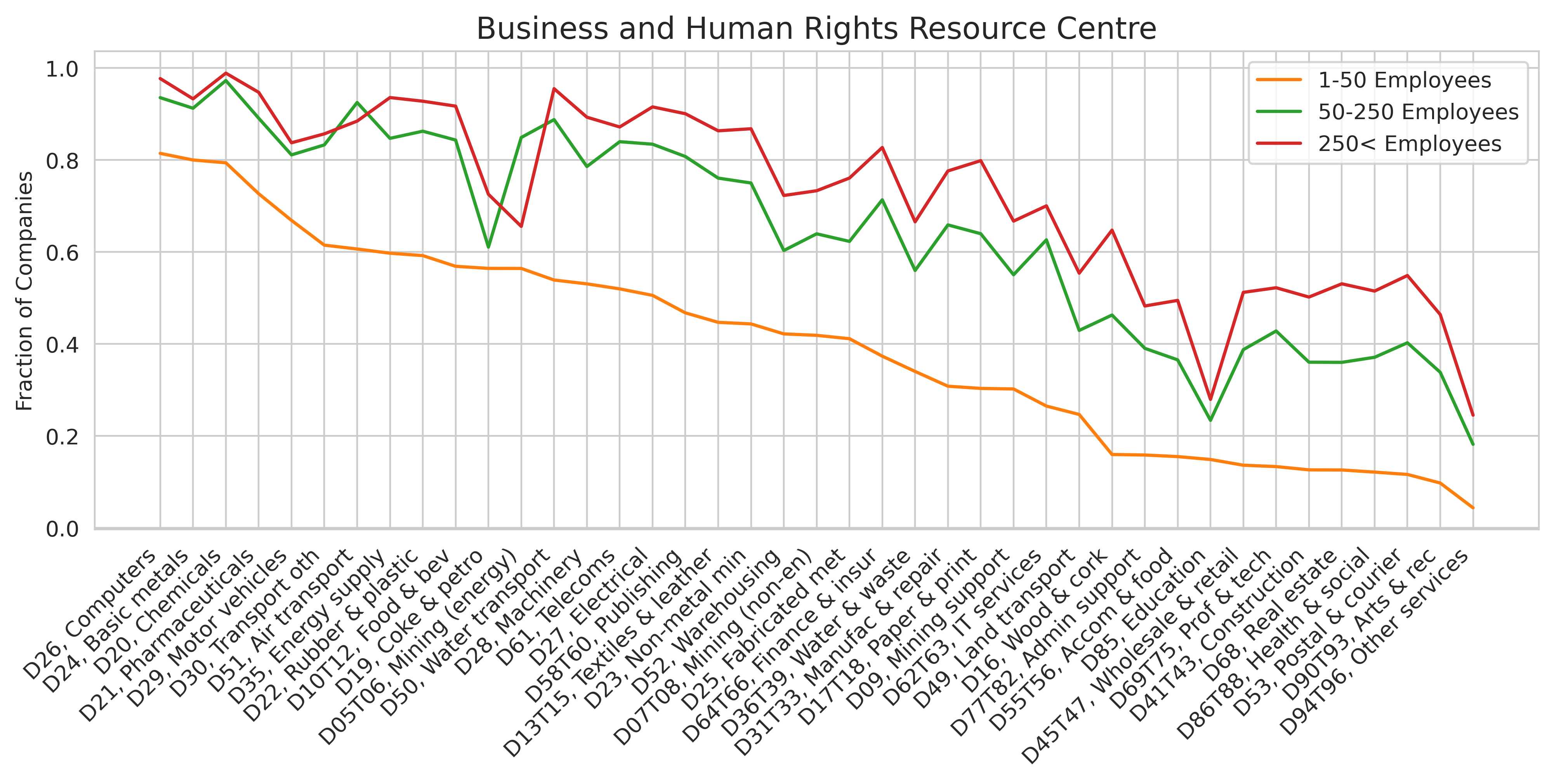}
    \caption{Results for the SCDD Risk Indicator by sector on tier one, stratified by firm size categories for (a) child and forced labour allegations and (b) HR-related lawsuits.}
    \label{fig:size_sec}
\end{figure}

\end{document}